# Uso de la inteligencia de ciberamenazas como apoyo a la comprensión del adversario aplicada al conflicto Rusia - Ucrania

*Using cyber threat intelligence to support adversary understanding applied to the Russia-Ukraine conflict*

**Oscar Sandoval Carlos**[1]

## Resumen

En organizaciones militares como el ejército, la Inteligencia de Ciberamenazas (CTI) apoya las operaciones cibernéticas proporcionando al comandante información esencial sobre el adversario, sus capacidades y objetivos mientras opera a través del ciberespacio. Este trabajo, combina la CTI con el marco MITRE ATT&CK para poder establecer un perfil de adversario. Además, se identifican las características de la fase de ataque mediante el análisis de la operación WhisperGate ocurrida en Ucrania en Enero del 2022, y finalmente se sugiere las medidas mínimas esenciales para la defensa.

## Abstract

In military organizations, Cyber Threat Intelligence (CTI) supports cyberspace operations by providing the commander with essential information about the adversary, their capabilities and objectives as they operate through cyberspace. This paper, combines CTI with the MITRE ATT&CK framework in order to establish an adversary profile. In addition, it identifies the characteristics of the attack phase by analyzing the WhisperGate operation that occurred in Ukraine in January 2022, and suggests the minimum essential measures for defense.

---

[1] Es Ingeniero Civil en Computación de la Universidad Arturo Prat, Postítulo en Gestión informática y Magister en Tecnologías de la Información y Gestión de la Pontificia Universidad Católica de Chile, Magister (c) en Ciberseguridad de la Universidad Adolfo Ibáñez y Certificate in Marketing de la Universidad de Berkeley EE.UU. Se desempeña como editor de publicaciones académicas en el CEEAG de la Academia de Guerra del ejército de Chile.

# Introducción

Todas las organizaciones están en riesgo de sufrir ataques cibernéticos de Rusia, advirtieron los gobiernos y las agencias de inteligencia de todo el mundo. Esta advertencia se produce en respuesta a la escalada de la invasión rusa en Ucrania ocurrido en febrero 2022, que incluye tácticas convencionales sobre el terreno, así como ciberataques[2].

Las operaciones cibernéticas rusas han apuntado a Ucrania con esfuerzos de desestabilización durante años, a través de ataques a la infraestructura crítica, operaciones de influencia, desconfiguración del sitio web, ataques contra bancos y redes militares ucranianas.

A medida que avanza el conflicto, aumenta la amenaza Rusa de una respuesta cibernética, debido a las múltiples sanciones impuestas por violar el derecho internacional. Según los gobiernos y las agencias de inteligencia de todo el mundo, "Rusia mantiene una serie de herramientas cibernéticas ofensivas que podría emplear contra las redes globales, desde denegaciones de servicio de bajo nivel hasta ataques destructivos dirigidos a infraestructura crítica". (MIT Technology Review, 2022).

Pero, *¿Qué es exactamente una ciberamenaza?*. Este artículo pretende contestar esta pregunta, desde una perspectiva de las ciberoperaciones militares, comenzando con una introducción a las intrusiones cibernéticas ocurridas en Ucrania para luego plantear las siguientes interrogantes que se van contestando a lo largo de las distintas secciones.

- *Qué sabemos sobre los diversos grupos de actividad cibernética que realizan operaciones en Ucrania y en el extranjero?*
- *¿Cómo se analiza un ciberataque y cuales son sus aspectos claves?*
- *¿Cómo protegerse y empezar a construir una ciberdefensa en base a las amenazas identificadas?*

Este trabajo hace uso de algunos de los resultados de investigación del Center For Threat Informed Defense de MITRE Corporation[3], a través del marco MITRE ATT&CK se utiliza la Cyber Threat Intelligence (CTI) disponible en distintas fuentes abiertas para identificar y comprender las características de una ciberamenaza, que pueden utilizarse para apoyar la identificación del adversario, la que puede tener como objetivo facilitar la planificación de la misión y determinar los mejores cursos de acción multidominio posibles.

---

[2] Fuente Cyber warfare in Ukraine poses a threat to the global system: https://www.ft.com/content/8e1e8176-2279-4596-9c0f-98629b4db5a6

[3] La misión del Centro es avanzar en el estado del arte y el estado de la práctica en la defensa informada sobre amenazas a nivel mundial: https://ctid.mitre-engenuity.org

Este documento también explora el análisis de intrusiones, a través de los modelos Cyber Kill Chain y Diamond Model, que tienen como objetivo detectar patrones de ataque de los adversarios para predecir futuros ciberataques.

Para finalizar se estudia el caso WhisperGate ocurrido en Enero del 2022 en Ucrania, examinando dos marcos complementarios: MITRE ATT&CK y MITRE D3FEND. Estos marcos describen las técnicas de los adversarios y las contramedidas de defensa respectivamente.

## Una mirada al conflicto desde el ciberespacio

El ciberespacio es considerado como el quinto dominio o dimensión de las operaciones de combate[4]. En este contexto, las ciberoperaciones son acciones militares que se desarrollan en el ciberespacio con los mismos objetivos que se producen en las otras cuatro dimensiones clásicas.

Las acciones en el ciberespacio se pueden dividir en dos tipos: Operaciones de Ciberdefensa y Operaciones de Ciberataques. En el campo de la ciberdefensa se busca como objetivo mantener la seguridad mediante la confidencialidad, integridad y disponibilidad de la información. Por otra parte, el ciberataque puede tener como objetivos la extracción o modificación de los datos del adversario y la denegación de sus servicios (DDoS). (Castro, 2021).

Figura 1. Taxonomía de las ciberoperaciones

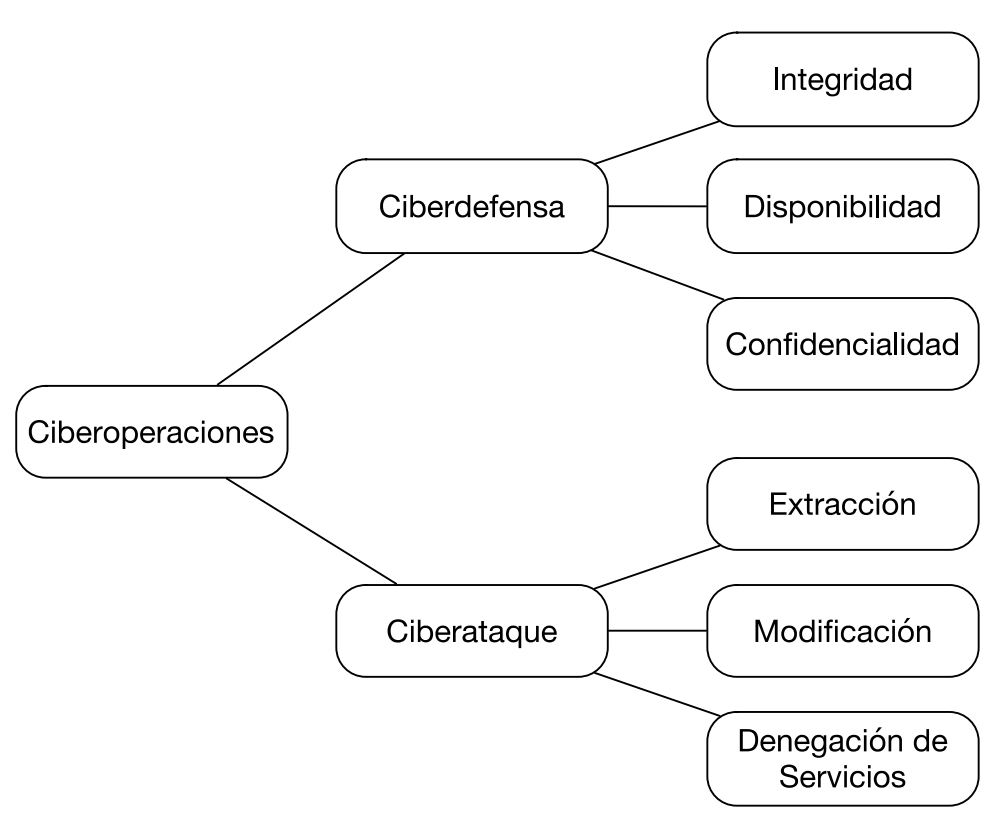

Fuente: Elaboración propia

[4] Los cuatro dominios tradicionales de las operaciones militares son tierra, mar, aire y espacio.

Las ciberoperaciones también pueden clasificarse según su nivel de ejecución[5]: ciberoperaciones estratégicas y ciberoperaciones operacionales y tácticas.

El principal objetivo de las operaciones de ciberataque estratégico es interrumpir la infraestructura crítica[6] de una nación y puede consistir en la extracción o modificación de valiosa información empresarial, tecnológica, militar, o en un ataque de denegación de servicio que paralice las operaciones, causando graves daños a la población civil. (Castro, 2021)

Según CERT-UA[7] El ataque fallido ocurrido en Abril 2022 por parte del grupo APT Sandworm, tenía como objetivo interrumpir el suministro eléctrico en ucrania, según diversas fuentes de inteligencia, la infección se produjo probablemente en febrero 2022 y se activó el ataque en Abril del mismo año, después de que el plan de invasión inicial ruso fallara. Lo anterior, indica que la decisión de no atacar la red eléctrica ucraniana fue una decisión estratégica y no formaba parte de los objetivos de la primera fase de invasión rusa, una razón es debido a que en una hipotética ocupación rusa, los ataques a la infraestructura hubieran agudizado el antagonismo de la población civil contra los ocupantes. El gran delta de tiempo entre la infección inicial y la fecha de ataque programada, fue probablemente la razón del por qué este ataque fallara, ya que proporcionó tiempo suficiente para que la ciberdefensa ucraniana coordinaran y ejecutaran acciones de remediación. (New York Times, 2022)

Por otro lado, el objetivo de las Operaciones de Ciberataque operacional y táctico es interrumpir los sistemas C4ISR[8] de una unidad militar, así como las redes de entidades gubernamentales y civiles que apoyan una operación de combate. Estas operaciones se llevan a cabo generalmente durante una guerra convencional y su objetivo es la extracción, modificación o denegación de servicio de información táctica que pueda afectar a los resultados de una batalla. (helpnet security, 2022)

Un ejemplo, de un ciberataque del tipo operacional ocurrió una semana antes del ataque Ruso a territorio Ucraniano a fines de febrero del 2022, donde varios bancos y sitios del

---

[5] La teoría militar divide la guerra en tres niveles: estratégico, operativo y táctico. Estos niveles están interrelacionados y lo que ocurre en un nivel influye en los demás. El nivel estratégico se ocupa de "cómo ganar una guerra". El nivel operativo, suele ocuparse de la realización de campañas y de cómo emplear las fuerzas en diversos teatros de operaciones. El nivel táctico es el ámbito de los enfrentamientos entre combatientes individuales y unidades en una situación de combate.

[6] Se define como Infraestructura Crítica (IC) a las instalaciones, sistema o parte de éste, que es esencial para el mantenimiento de las funciones sociales básicas, y cuya perturbación o destrucción, afectaría gravemente la salud, la integridad física, la seguridad y el bienestar social y económico de la población

[7] Significa Equipo de Respuesta ante Incidentes de Seguridad Informática (CSIRT) Ucrania : http://cert.gov.ua

[8] C4ISR: sistemas de mando y control, comunicaciones, informática, inteligencia, vigilancia y reconocimiento

gobierno Ucraniano fueron víctimas de ataques distribuidos de denegación de servicios (DDoS[9]). También se observaron ataques con el uso de mensajes SMS informando sobre la falla de los cajeros automáticos o disparando alarmas de bomba, y los soldados recibieron mensajes destinados a disuadirlos de luchar. La inteligencia norteamericana rápidamente culpó a la dirección de Inteligencia Militar Rusa (GRU) como autora de los ataques, lo que más tarde se interpretó como un intento de desestabilización para la invasión y sentar las bases para ciberataques más disruptivos contra Ucrania. (CHALFANT, 2022)

La figura siguiente muestra un resumen de los ciberataques recientes más importante contra ucrania

Figura 2: Línea de tiempo ciberataques en Ucrania

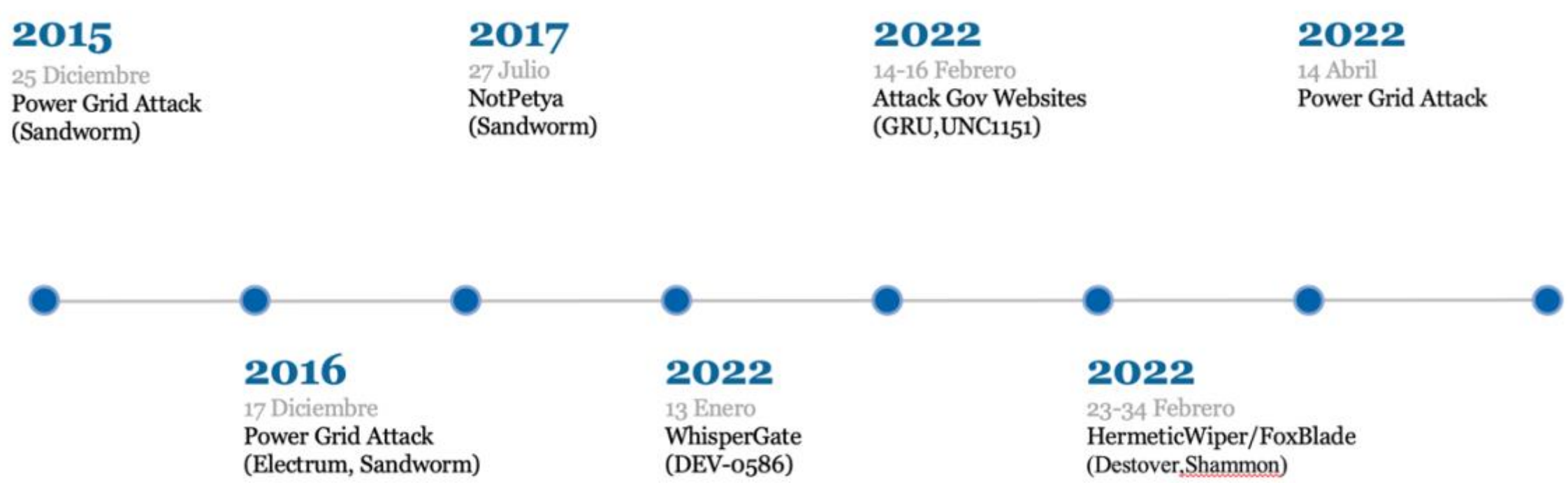

Fuente: Elaboración propia

Los grupos de actividad asociados a rusia son considerados del tipo APT[10], debido a su alta preparación y sofisticación en los ataques, utilizando complejas técnicas para realizar intrusiones en un sistema objetivo.

Para empezar a comprender las operación cibernéticas ocurridas en Ucrania, resulta imprescindible la identificación y posterior análisis de los grupos de actividad involucrados en el conflicto. A pesar de los avances de los software de detección de vulnerabilidades, los actores APT han demostrado continuamente la capacidad de comprometer los sistemas mediante el uso de herramientas avanzadas, malware personalizado y Exploit de "día

[9] Es un ataque a un sistema de computadoras o red que causa que un servicio o recurso sea inaccesible a los usuarios legítimos

[10] Una amenaza persistente avanzada, también conocida por sus siglas en inglés, APT, es un conjunto de procesos informáticos sigilosos orquestados por un tercero con la intención y la capacidad de atacar de forma avanzada y continuada en el tiempo, un objetivo determinado.

cero[11]" que los antivirus y los parches no pueden detectar o mitigar. Las respuestas a las intrusiones de APT requieren una evolución en el análisis para anticipar y mitigar futuras intrusiones basándose en el conocimiento de la amenaza. (Hutchins E. , 2011)

A continuación se propone un enfoque basado en la inteligencia de ciberamenazas para estudiar las intrusiones desde la perspectiva de los adversarios, donde cada fase de una intrusión se asigna a cursos de acción para la detección, mitigación y respuesta. La figura siguiente resume el enfoque propuesto en este trabajo.

Figura 3: Modelo propuesto para la defensa basada en inteligencia y en el análisis de las campañas de los adversarios

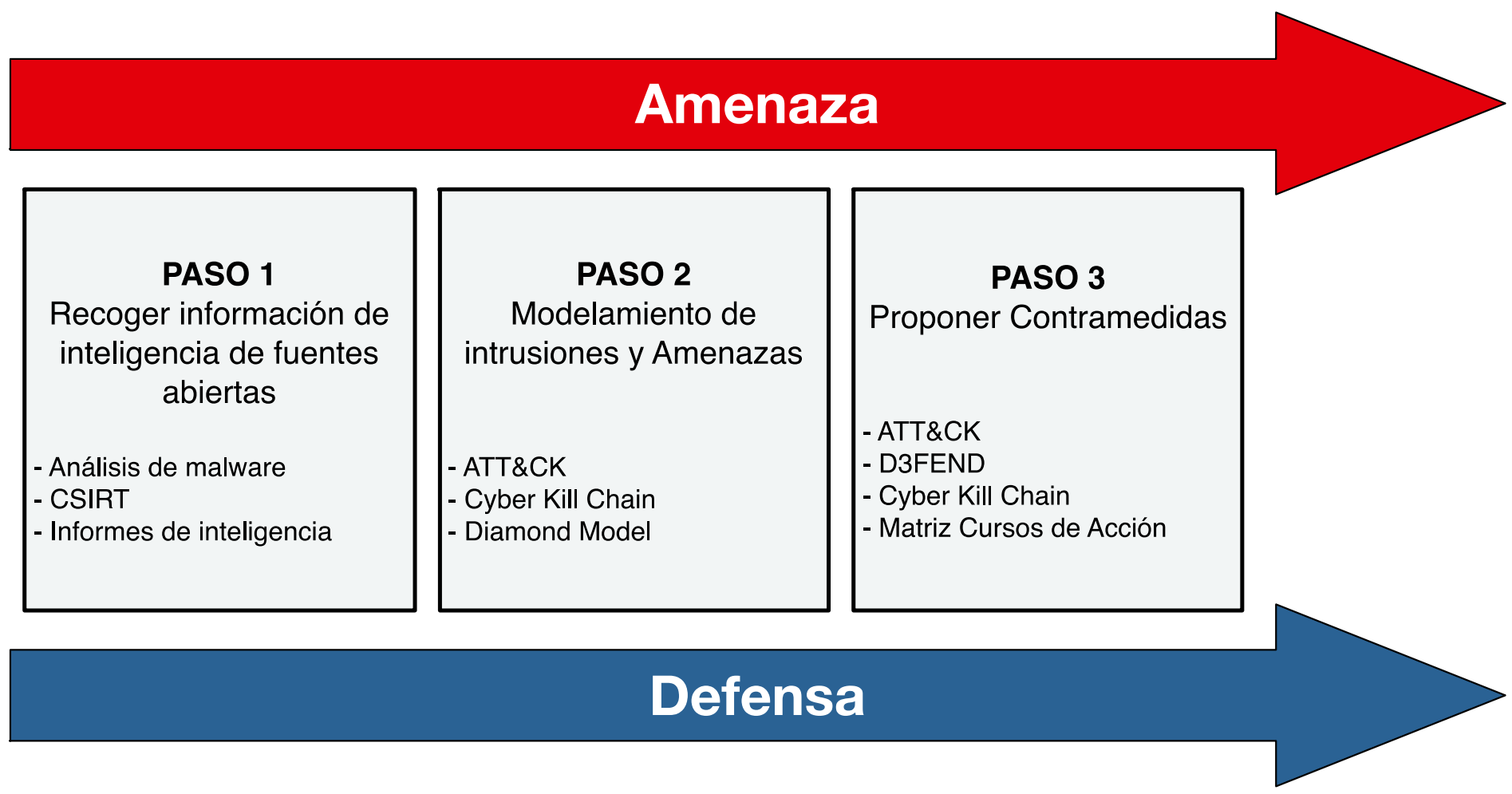

Fuente: Elaboración Propia

## Inteligencia de ciberamenazas: Detección de Tácticas, Técnicas y Procedimientos de adversarios

La inteligencia de ciberamenazas es el proceso de recopilación y análisis de datos relacionados con los ataques para rastrear, localizar, identificar y predecir las intenciones y actividades de los adversarios. (Lisa Institute, 2020). En esta sección se extraen las TTPs[12] de la Inteligencia de Amenazas adquirida mediante distintas fuentes abiertas y de uso público e introduciremos el marco ATT&CK, el estándar de la industria para identificar y documentar las TTPs comunes de los adversarios.

[11] Un Ataque de día cero es un ataque contra una aplicación o sistema que tiene como objetivo la ejecución de código malicioso gracias al conocimiento de vulnerabilidades que son desconocidas para los usuarios y para el fabricante del producto.
[12] Los TTP (Tácticas, Técnicas y Procedimientos) de un atacante son las acciones que realiza y métodos que utiliza a medida que avanza y desarrolla el ataque.

**Marco MITRE ATT&CK**

MITRE ATT&CK es una base de datos de conocimientos sobre amenazas a la seguridad que analiza cómo invade y se propaga un adversario a través de los sistemas informáticos, centrándose en los datos identificados en ataques reales. Los datos se describen mediante los tres elementos de la TTP:

- **Tácticas:** Los métodos y objetivos utilizados en el ataque
- **Técnicas:** Se han identificado 185 técnicas y 367 subtécnicas como métodos técnicos de ciberataque para alcanzar los objetivos tácticos
- **Procedimientos:** Los procesos secuenciales de un ataques.

Existen matrices ATT&CK sobre el comportamiento de los adversarios dirigidos a entornos empresariales, móviles y sistemas de control industrial. Además, se dispone de información relativa al software que utilizan los adversarios, técnicas de mitigación, ejemplos de procedimientos y recomendaciones de detección y recomendaciones de detección. La siguiente tabla muestra las tácticas de ATT&CK y sus definiciones.

| Táctica | Descripción |
|---|---|
| *Reconocimiento* | El adversario intenta reunir información que pueda utilizar para planificar futuras operaciones. |
| *Desarrollo de recursos* | El adversario intenta establecer los recursos que puede utilizar para apoyar las operaciones. |
| *Acceso inicial* | El adversario intenta acceder a la red objetivo. |
| *Ejecución* | El adversario intenta ejecutar un código malicioso. |
| *Persistencia* | El adversario intenta mantener su progreso. |
| *Elevación de privilegios* | El adversario intenta obtener permisos de nivel superior. |
| *Evasión de defensas* | El adversario intenta evitar la detección. |
| *Acceso a las credenciales* | El adversario intenta robar nombres de cuentas y contraseñas. |
| *Descubrimientos* | El adversario intenta determinar el entorno del objetivo. |
| *Movimiento lateral* | El adversario intenta moverse por el entorno del objetivo. |
| *Colección* | El adversario intenta reunir datos de interés para su objetivo. |
| *Mando y Control* | El adversario intenta comunicarse con los sistemas comprometidos para controlarlos. |
| *Exfiltración* | El adversario intenta robar datos. |
| *Impacto* | El adversario intenta manipular, interrumpir o destruir los sistemas y datos del objetivo. |

Tabla 1: MITRE ATT&CK Tácticas

**Uso de Inteligencia de ciberamenazas para la creación del perfil de adversario**

Según la información entregada por la agencia de ciberseguridad norteamericana[13] existen varios grupos de actividad asociados a Rusia, entre los más importantes podemos nombrar:

- APT28 es un grupo de amenazas que se ha atribuido al Centro Principal de Servicios Especiales de la Dirección Principal de Inteligencia del Estado Mayor de Rusia (GRU)
- APT29 es un grupo de amenazas que se ha atribuido al Servicio de Inteligencia Exterior (SVR) de Rusia. Han operado desde al menos 2008.
- Turla es un grupo de amenazas con sede en Rusia que ha infectado a víctimas en más de 45 países, abarcando una amplia gama de industrias.

Figura 4: Grupos de amenazas asociados con Rusia

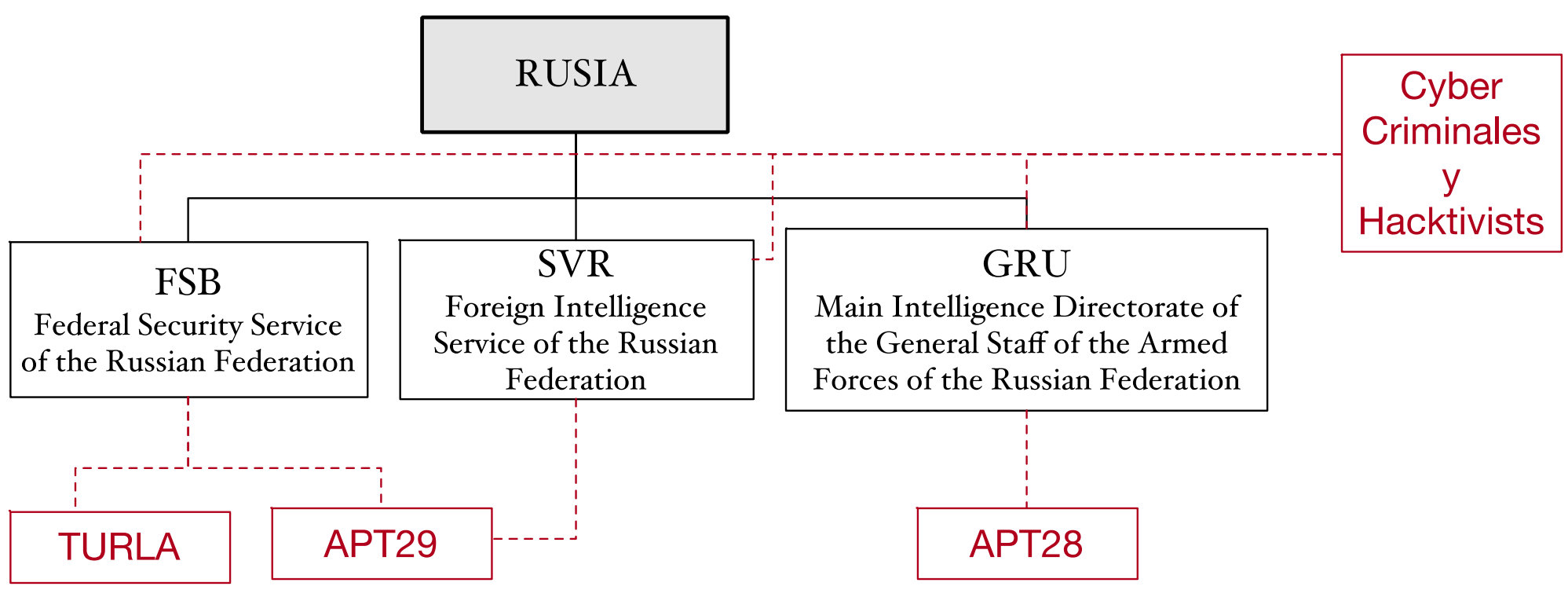

Fuente: Elaboración propia

Otros grupos con una participación activa en el conflicto son *Sandworm*, caracterizado principalmente por el uso del malware *BlackEnergy* y sus ataques a la red eléctrica Ucraniana y *Dev-0586* asociado a los ataques del tipo *Wiper*. (Cybersecurity & Infrastructure Security Agency, 2022).

La creación de un perfil del adversario es el primer paso para comprender las amenazas y elaborar un plan de alto nivel sobre cómo empezar a defenderse. Utilizando el marco MITRE ATT&CK se puede obtener información sobre grupos de actividad en el conflicto, las tablas siguientes muestran el perfil de adversario de los grupos APT asociados a Rusia.

[13] https://www.cisa.gov

| | |
|---|---|
| *Descripción:* | **APT28** es un grupo de amenazas que se ha atribuido al Centro Principal de Servicios Especiales de la Dirección Principal de Inteligencia del Estado Mayor de Rusia (GRU) |
| *Otros nombres asociados:* | SNAKEMACKEREL, Swallowtail, Group 74, Sednit, Sofacy, Pawn Storm, Fancy Bear, STRONTIUM, Tsar Team, Threat Group-4127. |
| *Técnicas:* | • Access Token Manipulation<br>• Account Manipulation<br>• Acquire Infrastructure<br>• Archive Collected Data<br>• Automated Collection<br>• Active Scanning<br>• Application Layer Protocol<br>• Boot or Logon Autostart Execution<br>• y más |
| *Software:* | ADVSTORESHELL, Cannon, certutil, CHOPSTICK, CORESHELL, DealersChoice, Downdelph, Drovorub, Forfiles, Fysbis, y más |

Tabla 2: Perfil de adversario APT28

| | |
|---|---|
| *Descripción:* | **APT29** es un grupo de amenazas que se ha atribuido al Servicio de Inteligencia Exterior (SVR) de Rusia. Han operado desde al menos 2008. |
| *Otros nombres asociados:* | NobleBaron, Dark Halo, StellarParticle, NOBELIUM, UNC2452, YTTRIUM, The Dukes, Cozy Bear, CozyDuke |
| *Técnicas:* | • Abuse Elevation Control Mechanism<br>• Account Discovery<br>• Account Manipulation<br>• Exploit Public-Facing Application<br>• Exploitation for Client Execution<br>• External Remote Services<br>• Forge Web Credentials<br>• Ingress Tool Transfer<br>• Y más |
| *Software:* | AdFind, BoomBox, CloudDuke, Cobalt Strike, CozyCar, EnvyScout, FatDuke, GoldFinder, GoldMax, HAMMERTOSS, Mimikatz, NativeZone, y más |

Tabla 3: Perfil de adversario APT29

| | |
|---|---|
| *Descripción:* | **Turla** es un grupo de amenazas con sede en Rusia que ha infectado a víctimas en más de 45 países, abarcando una amplia gama de industrias, incluyendo el gobierno, las embajadas, el ejército, la educación, la investigación y las empresas farmacéuticas desde 2004. |
| *Otros nombres asociados:* | Group 88, Belugasturgeon, Waterbug, WhiteBear, VENOMOUS BEAR, Snake, Krypton |
| *Técnicas:* | • Exploitation for Privilege Escalation<br>• File and Directory Discovery<br>• Group Policy Discovery<br>• Ingress Tool Transfer<br>• Password Policy Discovery<br>• Process Injection<br>• Remote System Discovery<br>• System Network Configuration Discovery<br>• System Service Discovery<br>• Web Service |
| *Software:* | Arp, Carbon, certutil, ComRAT, Crutch, Empire, Epic, Gazer, HyperStack, Kazuar, LightNeuron, Mimikatz, y más |

Tabla 4: Perfil de adversario Turla

## Uso de ATT&CK para la extracción de tácticas, técnicas y procedimientos

MITRE ha desarrollado el Navegador ATT&CK[14], una aplicación web que representa las técnicas de forma dinámica. Se puede utilizar para seleccionar técnicas específicas basadas en un grupo de amenazas. La figura siguiente muestra la matriz ATT&CK identificando en color rojo las TTPs asociadas al grupo APT28.

Figura 5: Tácticas y Técnicas utilizadas por el grupo APT28

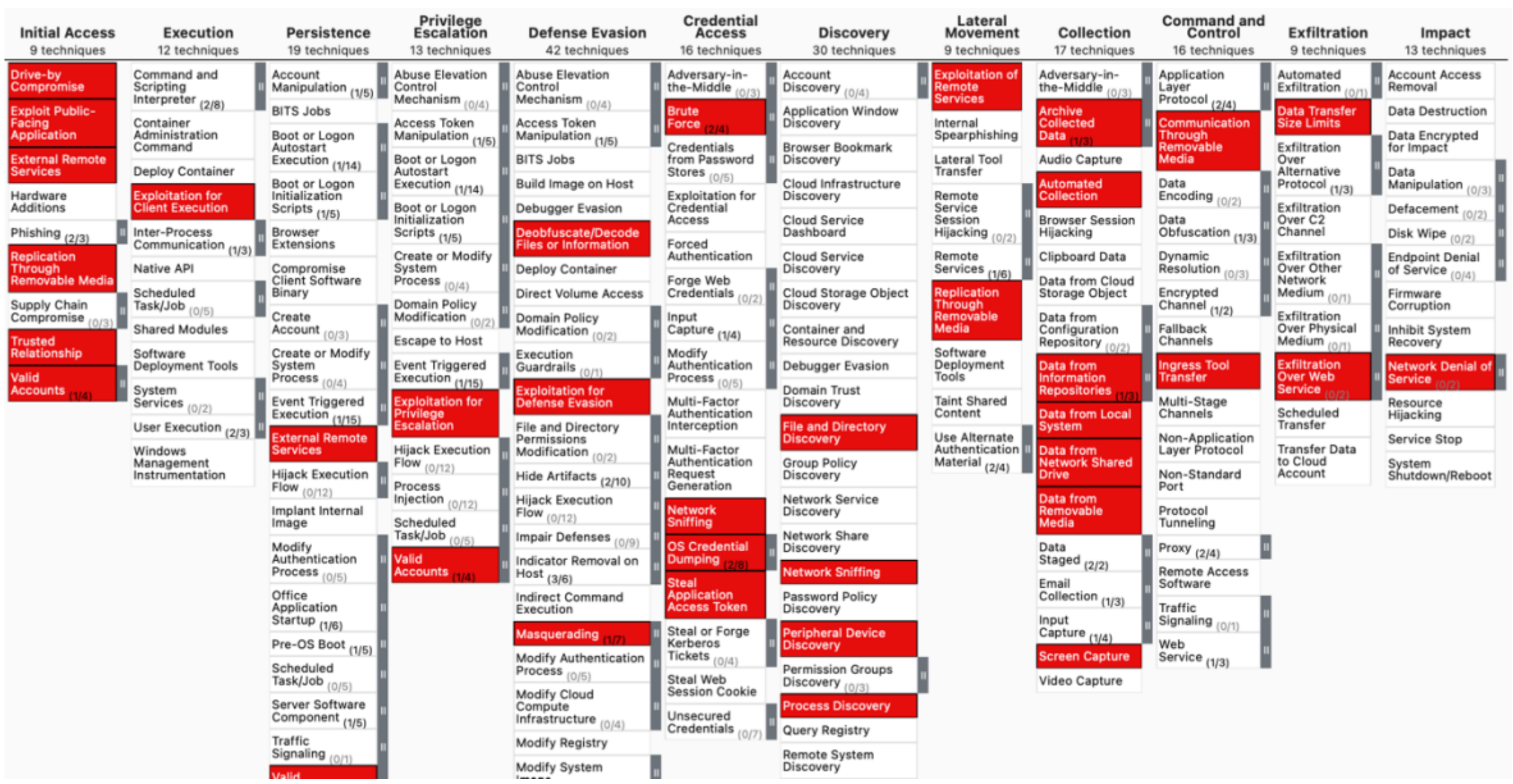

Fuente: https://mitre-attack.github.io/attack-navigator/

La matriz ATT&CK visualiza y describe la relación entre tácticas y técnicas, como se muestra en la figura 5. Las tácticas representan el motivo del comportamiento de los atacantes, mientras que las técnicas describen la forma de lograr y obtener el objetivo táctico.

De la matriz de la figura 5, se extrae manualmente la información necesaria para construir la tabla 5. Se considera cada técnica como un paso de ataque que puede ser realizado por un atacante. La siguiente tabla muestra las técnicas identificadas a través del marco ATT&CK para el grupo APT28.

[14] https://mitre-attack.github.io/attack-navigator/

| Táctica | Técnica |
|---|---|
| *Acceso inicial* | T1190 – Exploit Public-Facing Application<br>T1133 – External Remote Services<br>T1091 – Replication Through Removable Media<br>T1199 – Trusted Relationship<br>T1078 – Valid Accounts |
| *Ejecución* | T1203 – Exploitation for Client Execution |
| *Persistencia* | T1133 - External Remote Services<br>T1078 – Valid Account |
| *Escalamiento de privilegios* | T1068 – Exploitation for Privilege Escalation<br>T1078 – Valid Accounts |
| *Evasión de defensas* | T1140 – Deobfuscate/Decode Files or Information<br>T1211 – Exploitation for Defense Evasión<br>T1036 – Masquerading<br>T1027 – Obfuscated Files or Information<br>T1014 – RootKit<br>T1221 – Template Injection<br>T1078 – Valid Accounts |
| *Acceso credenciales* | T1110 – Brute Force<br>T1040 – Network Sniffing<br>T1003 – OS Credential Dumping<br>T1528 – Steal Application Access Token |
| *Descubrimiento* | T1083 – File and Directory Discovery<br>T1040 – Network Sniffing<br>T1120 – Peripheral Device Discovery<br>T1057 – Process Discovery |
| *Movimiento lateral* | T1210 – Exploitation of Remote Services<br>T1091 – Replication Through Removable Media |
| *Comando y Control* | T1092 – Communication Through Removable Media<br>T1105 – Ingress Tool Transfer |
| *Filtración* | T1030 - Data Transfer Size Limits<br>T1567 Exfiltration Over Web Service |
| *Impacto* | T1498 – Network Denial of Service |

Tabla 5: Tácticas y técnicas utilizados por la amenaza **APT28**

La identificación de las APT y sus TTPs asociadas, es el primer paso para la elaboración de una inteligencia de ciberamenazas para cualquier organización. Con la información anterior, ya es posible analizar y organizar la amenaza en un flujo técnico que permita formular escenarios de ataques. La figura siguiente se utiliza un escenario hipotético del grupo APT28 utilizando las TTPs identificadas para desplazarse por la red de una victima.

Figura 6: Diagrama de flujo escenario APT28

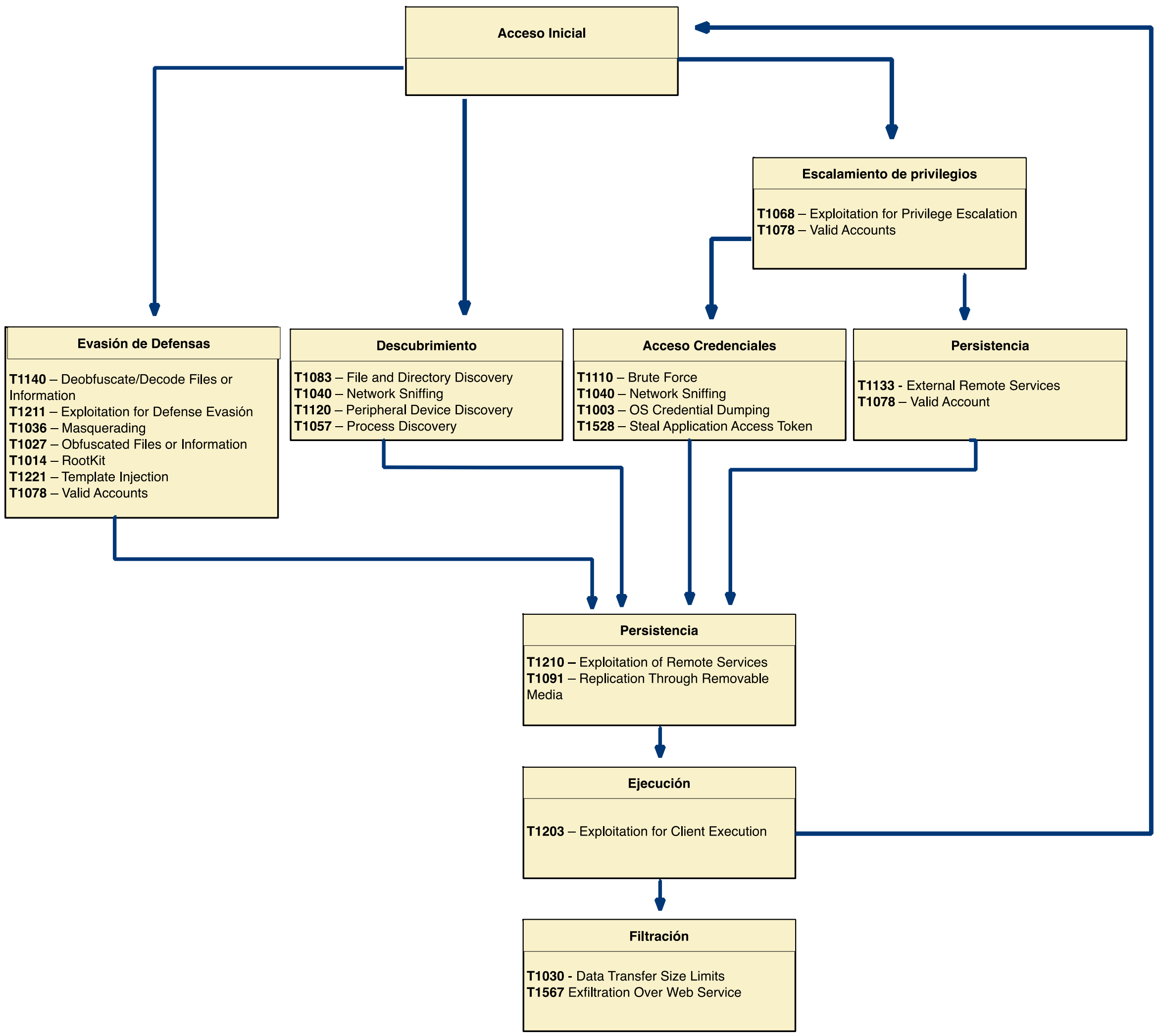

Fuente: Elaboración Propia

## Modelado de intrusiones : Análisis de ciberataques a través de la cadena de la muerte

Las técnicas de modelado de ciberataques tienen como objetivo detectar patrones de ataque de los adversarios para predecir futuros ciberataques, nos ayudan a caracterizar las amenazas detectadas, hacerles seguimiento y luego encontrar la manera de contrarrestarlas.

En esta sección, se analiza dos de los enfoques existentes más avanzados relacionados con las técnicas de modelado de ciberataques.

**Cadena de la Muerte**

Cyber Kill Chain, fue establecida el año 2011 por investigadores de la Lookheed Martin Corporation, se basa en la experiencia de los autores en la defensa de redes y en sus vínculos con el Departamento de Defensa de EE.UU, y es un método para modelar intrusiones en una red informática basada en la doctrina militar conocida como "Kill Chain". (Hutchins, 2011)

En este modelo, un agresor debe desarrollar un método para entrar en un entorno de confianza con el fin de establecer una presencia y tomar acciones hacia algún objetivo que le gustaría lograr en la red (por ejemplo, el robo de datos). En términos de una cadena de intrusión, se define como un proceso que contiene las siguientes fases:

- **Paso 1: Reconocimiento** - El atacante obtiene información sobre la estructura de la red, el personal y los activos que la componen.
- **Paso 2: Armamento** - El atacante utiliza un exploit para crear una actividad maliciosa que explotará una vulnerabilidad dentro de la red informática del objetivo.
- **Paso 3: Entrega** - El atacante envía el malware hacia la red del objetivo (por ejemplo, a través del correo electrónico o la web).
- **Paso 4: Explotación -** El exploit se ejecuta dentro de la red informática de la victima y el atacante obtiene acceso.
- **Paso 5: Instalación** - Instalación del malware dentro de la red informática de la víctima. Esta fase puede durar semanas o meses, ya que puede implicar ampliar el acceso, ganar persistencia y realizar un reconocimiento interno dentro de la red de la víctima.
- **Paso 6: Mando y control** - El atacante crea una comunicación con el malware (es decir, un canal de mando y control (C2C)) para mantener el acceso a los dispositivos comprometidos.
- **Paso 7: Acción sobre los objetivos** - El atacante realiza las tareas que necesita para lograr sus objetivos deseados en la red objetivo. Este paso puede llevar semanas o meses.

Cyber Kill Chain ofrece un método para entender las fases de un ataque cibernético y aplicar cursos de acción defensivos en toda la cadena de intrusión.

## Modelo de Diamante

El Modelo Diamante de Análisis de Intrusiones, describe cómo un adversario utiliza la capacidad de una infraestructura contra una víctima. Este modelo consta de cuatro elementos básicos: adversario, infraestructura, capacidad y víctima que proporcionan cuatro cuadrantes como un diamante. A medida que el adversario descubre información adicional sobre el ataque a través del análisis o de nueva información, se pueden rellenar nuevos vértices del diamante o refinar y actualizar los existentes. (Caltagirone, 2013)

El elemento básico del modelo es el denominado Evento, como se muestra en la figura siguiente, es la representación de la acción de un *Adversario* que, desplegando una determinada *Capacidad* sobre una *Infraestructura* concreta, ataca a una *Victima*.

Figura 6: Se ilustra el pivote analítico utilizando el Diamante. Una de las características más potentes del Diamante, el pivoteo, permite explotar la relación fundamental entre las características para descubrir nuevos conocimientos sobre una actividad maliciosa.

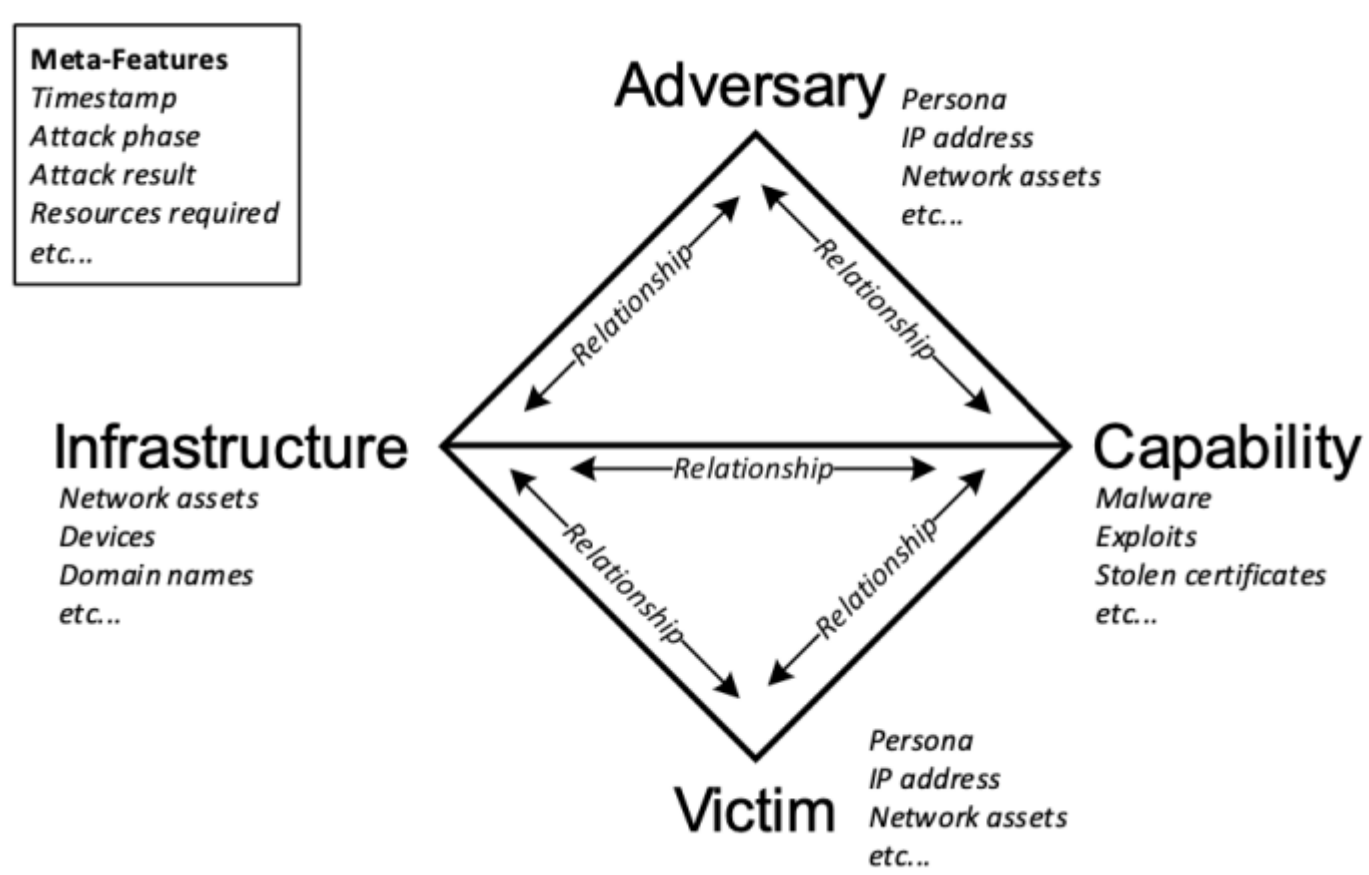

Fuente: Cook, A., Janicke, H., Smith, R., & Maglaras, L. (2017). The industrial control system cyber defence triage process. Computers & Security, 70, 467-481.

Los arcos, a su vez, está etiquetado con atributos que incluye si el camino es necesario u opcional (denotado por AND u OR respectivamente), o si se considera el arco como REAL o HIPÓTETICO y, finalmente, una descripción sobre la información o el recurso que el evento precedente proporciona y que se requiere para que el siguiente evento ocurra.

Un evento define tan sólo un paso de una serie de ellos que el adversario debe ejecutar para lograr su objetivo. Así, los eventos, ordenados en fases y en virtud de una relación concreta adversario-víctima, se configuran en forma de Hilos de Actividad, que representan el flujo de operaciones de un adversario. Ambos, Eventos e Hilos de Actividad constituyen elementos necesarios para lograr comprender la actividad de intrusión. (Caltagirone, 2013)

## Caso de estudio: Ataque WhisperGate contra Ucrania Enero 2022

Analizado por primera vez por Microsoft Threat Intelligence Center (MSTIC), WhisperGate se detectó el 13 de enero de 2022. Según el informe de MSTIC, este malware se lanzó explícitamente contra varias organizaciones ucranianas en ataques con motivos geopolíticos.

WhisperGate se disfraza de un malware tipo Ransomware pidiendo una nota de rescate para la recuperación de los datos robados, pero es un malware destructivo y está diseñado para hacer que los dispositivos infectados sean irrecuperables.

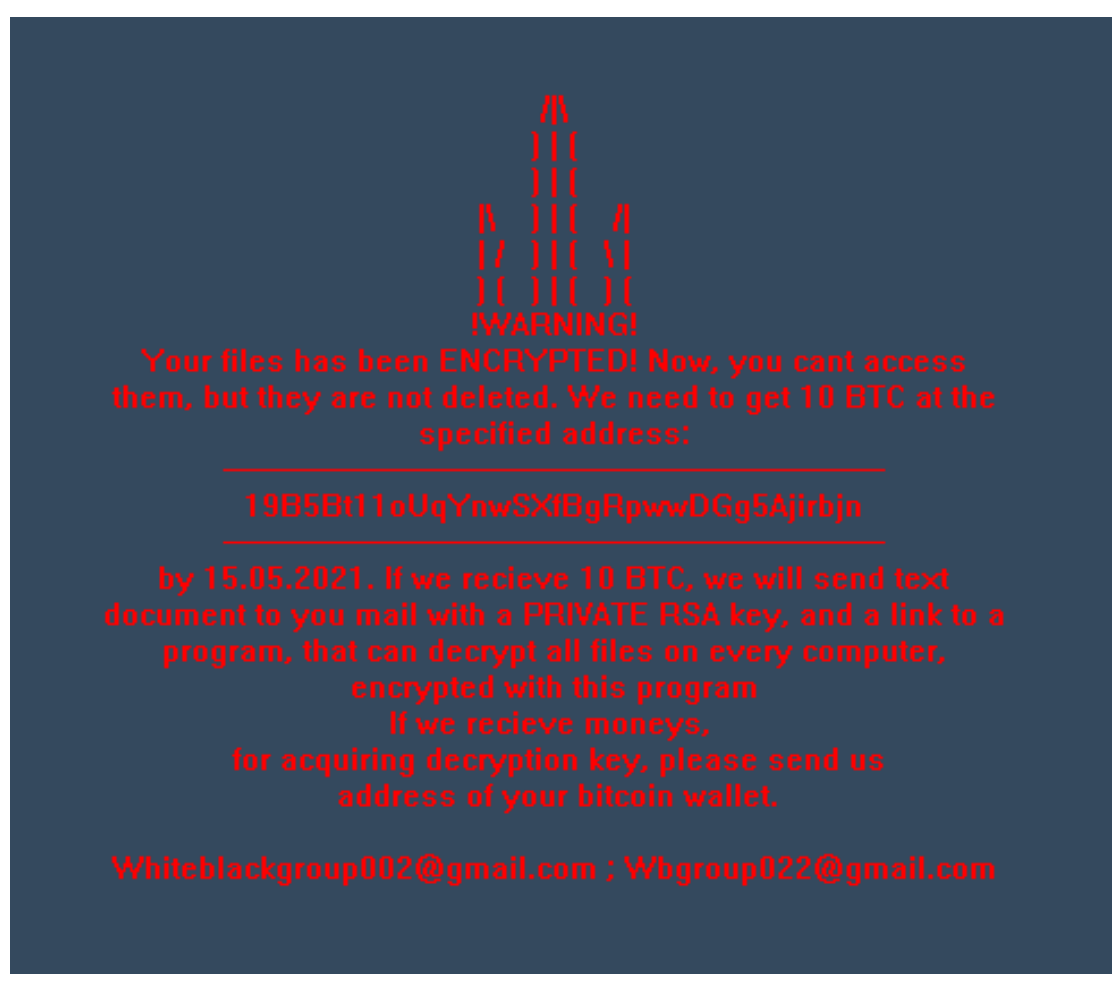

Figura 7: Nota de rescate WhisperGate, fuente: Talos Intelligence[15]

Microsoft está siguiendo estos ataques como DEV-0586. La designación "DEV" indica que se trata de "un nombre temporal dado a una actividad de amenaza desconocida, emergente o en desarrollo". (Cybersecurity & Infrastructure Security Agency, 2022)

Las Tablas siguiente revela el mapeo de las actividades de WhisperGate reportadas por la agencia de ciberseguridad de Estados Unidos (CISA)[16], con las tácticas y técnicas del marco MITRE ATT&CK.

---

[15] https://blog.talosintelligence.com/2022/01/ukraine-campaign-delivers-defacement.html

[16] https://www.cisa.gov/uscert/ncas/alerts/aa22-057a

| Categoría | Descripción | |
|---|---|---|
| *Acceso inicial* | • Sin Información | |
| *Ejecución* | • **T1059.003** Command and Scripting Interpreter: Windows Command Shell<br>• **T1059.001** Command and Scripting Interpreter: PowerShell | La primera etapa del malware WhisperGate utiliza un comando de Windows Command Shell para ejecutar el malware. Luego utiliza PowerShell para conectar su servidor de Mando y Control (C2) y descargar cargas útiles adicionales |
| *Persistencia* | • **T1542.003** Pre-OS Boot: Bootkit | WhisperGate modifica el Registro Maestro de Arranque (MBR). Dado que el MBR alterado es la primera sección del disco después de completar la inicialización del hardware por la BIOS, el malware evade la defensa. |
| *Escalamiento de privilegios* | • Sin información | |
| *Evasión de defensas* | • **T1027** Obfuscated Files or Information | WhisperGate entrega comandos PowerShell en Base64 |
| *Acceso credenciales* | • Sin información | |
| *Descubrimiento* | • **T1083** File and Directory Discovery | WhisperGate busca extensiones de archivo específicas en determinados directorios para alterar su contenido. |
| *Movimiento lateral* | • Sin información | |
| *Comando y Control* | • **T1105** Ingress Tool Transfer | WhisperGate descarga el archivo corruptor payload desde el canal de Discord alojado por el grupo APT. El enlace de descarga del ejecutable malicioso está codificado en el stage2.exe. |
| *Filtración* | • Sin Información | |
| *Impacto* | • **T1561** Disk Wipe<br>• **T1485** Data Destruction | WhisperGate sobrescribe el Master Boot. (Cuando se sobrescribe el MBR, el sistema infectado no arranca tras el apagado). Luego, WhisperGate sobrescribe los archivos y afecta negativamente a su integridad. Además, el malware cambia el nombre de los archivos para aumentar su impacto. |

Tabla 6: TTP usados por el grupo APT DEV-0586

La tabla anterior, utilizando la matriz ATT&CK, nos entrega información detallada sobre las tácticas y técnicas utilizadas por el grupo DEV-0586 en el ataque WhisperGate, pero no nos muestra el orden lógico del ataque, que permita identificar la secuencia y los elementos relacionados en la intrusión.

El análisis dirigido es una de las fortalezas del Modelo de Diamante. El análisis dirigido es la actividad que permite descubrir elementos relacionados capaces tanto de sustentar una hipótesis previa del ataque como la de generar nuevas, permitiendo analizar la intrusión en términos de capacidad del atacante y la infraestructura utilizada. La figura siguiente muestra un análisis dirigido del ataque WhisperGate en base a los datos recopilados en la tabla 6.

figura 8 se puede observar al *Adversario* hacer uso de su *Capacidad* para utilizar distintos vectores de ataque y así obtener acceso a los sistemas, luego utiliza la *Infraestructura* de la victimas para ejecutar el malware y descargar el software de comando y control, finalmente utiliza su *Capacidad* contra la *Victima* para ejecutar el ataque.

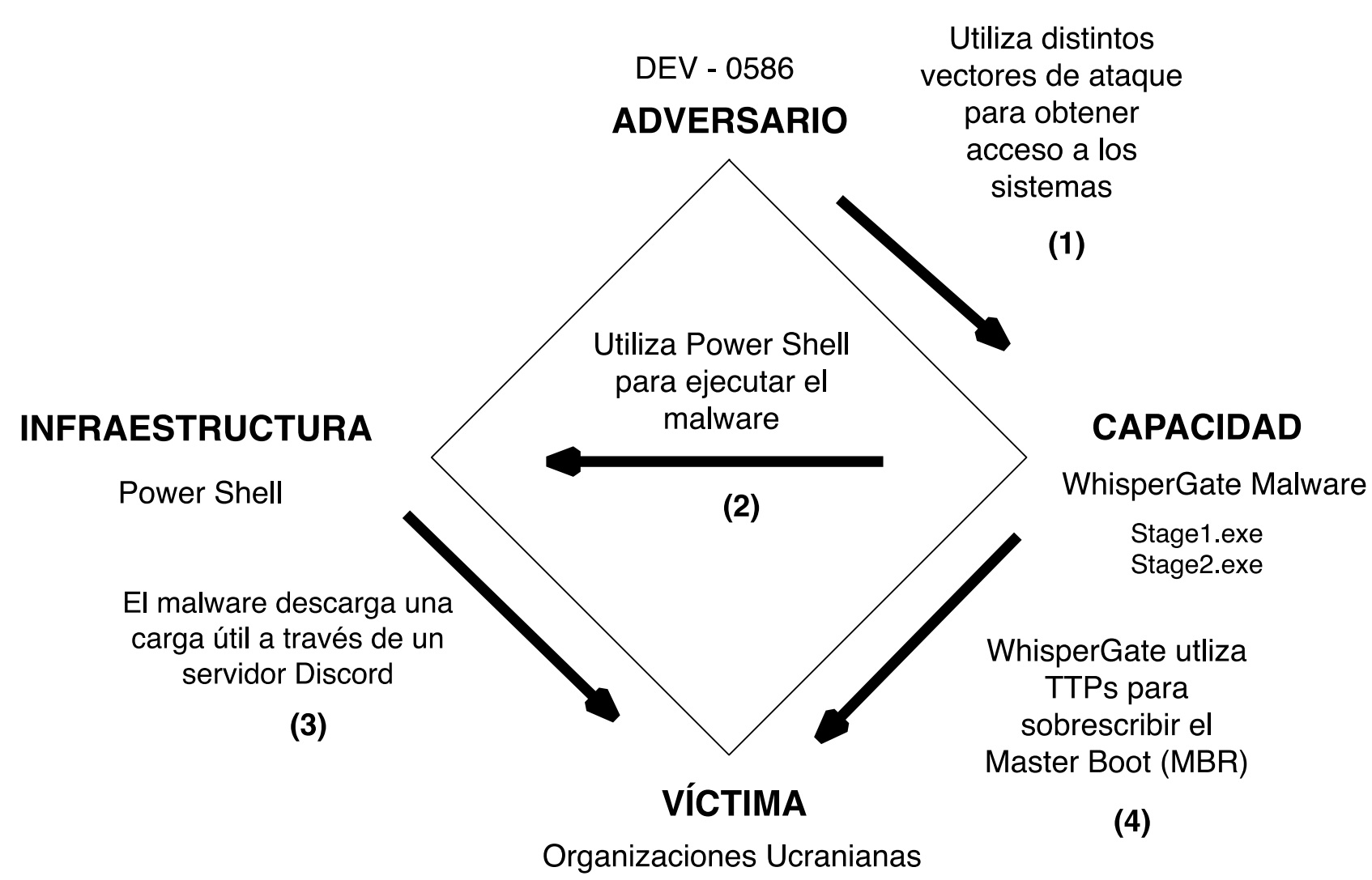

Fuente: Elaboración propia

La figura anterior nos muestra una secuencia lógica de la intrusión, descubriendo las capacidades del adversario y la infraestructura sobre las que son desplegadas. Para ampliar este análisis, se utiliza el marco Cyber Kill Chain para estructurar el ataque en siete pasos secuenciales. La tabla siguiente describe la secuencia del ataque WhisperGate en términos de la cadena de la muerte.

| **Fase** | **Descripción** |
| --- | --- |
| Reconocimiento | Aunque no ha sido reportado oficialmente, dada la gran cantidad de eventos a distintas organizaciones, se sugiere un ataque oportunista frente a uno dirigido. |
| Armamento | Una vez completado el reconocimiento, El adversario prepara un archivo ejecutable malicioso codificado como stage1.exe o stage2.exe. |
| Entrega | se presume que se utilizaron contraseñas robadas en campañas anteriores para tener acceso a los sistemas. Al mismo tiempo, no se descartan otros dos posibles vectores de ataque, como la explotación de las vulnerabilidades OctoberCMS y Log4j. |
| Explotación | WhisperGate utiliza un comando de Windows Command Shell para ejecutar el malware y se ejecuta a través de Impacket, una capacidad disponible públicamente utilizada a menudo por los actores de amenazas para el movimiento lateral y la ejecución. |
| Instalación | WhisperGate modifica el Registro Maestro de Arranque (MBR). Dado que el MBR alterado es la primera sección del disco después de completar la inicialización del hardware por la BIOS, el malware evade la defensa. El malware también busca extensiones de archivo específicas en determinados directorios para alterar su contenido. |
| Mando y Control | la carga útil es descargada a través de un archivo DLL malicioso alojado en un servidor Discord, que suelta y ejecuta otra carga útil con el objetivo de destruir los archivos de las máquinas infectadas. |
| Acción sobre los objetivos | La primera carga útil de esta infección es responsable del intento inicial de limpiar los sistemas. El ejecutable de malware borra el registro maestro de arranque (MBR) y lo reemplaza por el código responsable de mostrar la nota de rescate, pero WhisperGate no pretende ser un intento de rescate real, ya que el MBR está completamente sobrescrito y no tiene opciones de recuperación. El malware también intenta destruir la partición C:\ sobrescribiéndola con datos fijos.<br><br>También localiza archivos en ciertos directorios del sistema y sobrescribe el contenido del archivo con un número fijo de 0xCC bytes |
| | (tamaño total del archivo de 1 MB). Después de sobrescribir el contenido, el malware cambia el nombre de cada archivo con una extensión de cuatro bytes aparentemente aleatoria. |

Tabla 7: Modelo de intrusión Kill Chain WhisperGate

La tabla anterior ilustra claramente el marco de un ciberataque exitoso, clasificando las acciones tomadas por el ataque WhisperGate en etapas secuenciales, desde la recopilación de la información sobre un objetivo, pasando por la preparación de todas las herramientas para ejecutar el malware, hasta completar finalmente el objetivo del ataque.

La cadena Cyber Kill Chain se puede combinar con el modelo de diamante para proporcionar una representación bidimensional de las relaciones entre los eventos generados por el *Adversario* en las diferentes fases de la Cyber Kill Chain.

Figura 9: Modelo de Diamante Ataque WhisperGate

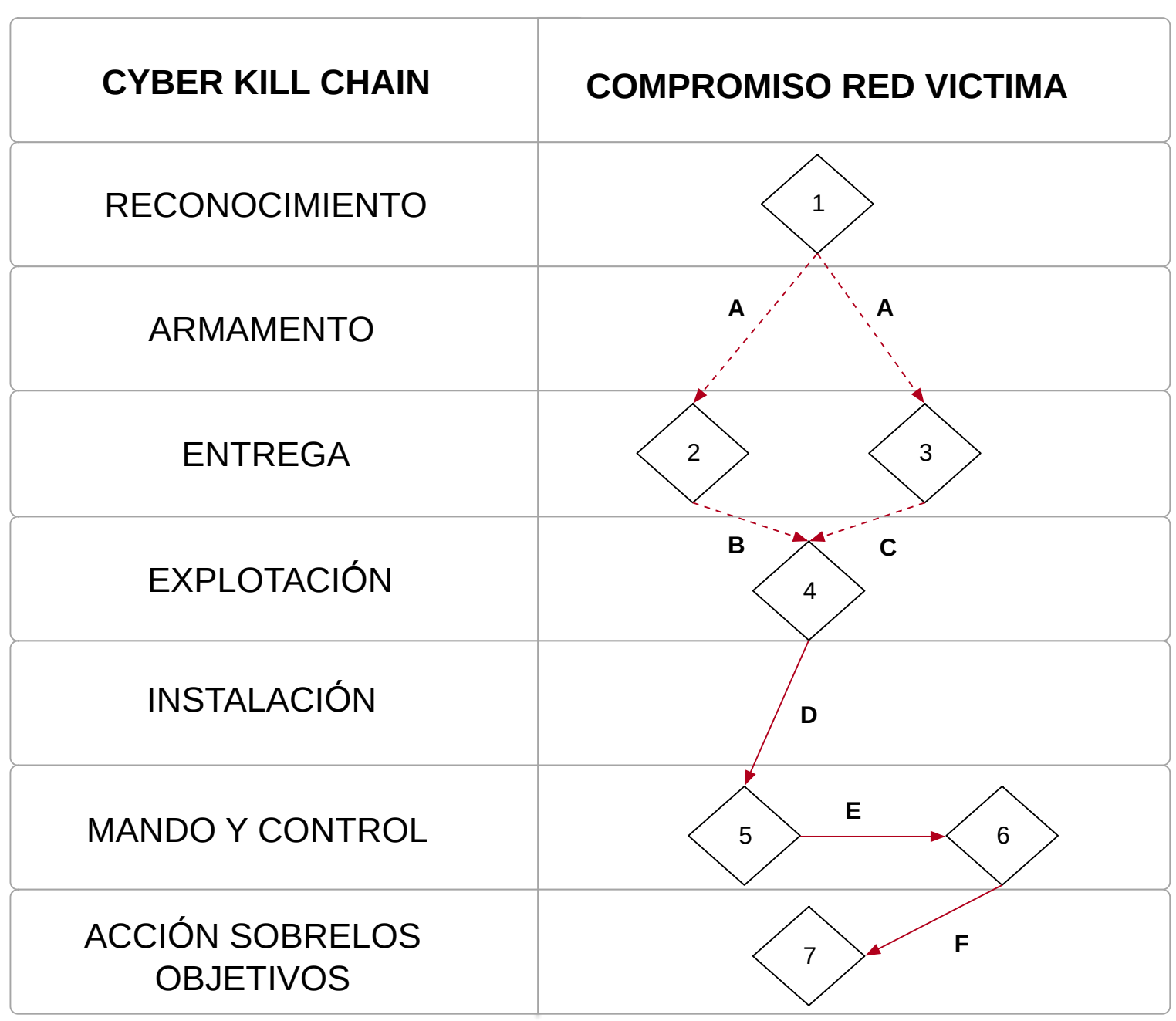

Fuente: Elaboración propia

| Evento | Hipótesis / Real | Descripción |
|---|---|---|
| **1** | Hipótesis | El Adversario utiliza la inteligencia de fuente abierta para seleccionar al adversario. |
| **2** | Hipótesis | El adversario utiliza contraseña robadas en campañas anteriores para obtener acceso a los sistemas. |
| **3** | Hipótesis | El adversario explota las vulnerabilidades OctoberCMS y Log4j para obtener acceso a los sistemas. |
| **4** | Real | El adversario utiliza Windows Command Shell para ejecutar el malware |
| **5** | Real | El malware descarga una carga útil a través de un archivo DLL malicioso alojado en un servidor Discord, |
| **6** | Real | Se descarga una segunda carga útil con el objetivo de destruir los archivos de las máquinas infectadas. |
| **7** | Real | El malware borra el registro maestro de arranque (MBR) y lo reemplaza por el código responsable de mostrar la nota de rescate. También el malware cambia el nombre de cada archivo con una extensión de cuatro bytes aparentemente aleatoria. |

Tabla 8: Eventos del hilo de actividad en el ataque WhisperGate

| Arco | Confianza | And/OR | Hipótesis /Real | Proporciona |
|---|---|---|---|---|
| **A** | Baja | OR | Hipótesis | Proporciona información sobre víctima |
| **B** | Baja | OR | Hipótesis | Proporciona acceso a la red de la víctima |
| **C** | Baja | OR | Hipótesis | Proporciona acceso a la red de la víctima |
| **D** | Alta | AND | Real | Proporciona la ejecución del Malware |
| **E** | Alta | AND | Real | Proporciona el establecimiento de la conexión remota. |
| **F** | Alta | AND | Real | Proporciona la carga útil del malware |

Tabla 9: Hilos de actividad para la figura 9.

En la tabla 8, cada evento se describe como real o hipotético. Un evento real significa que hay pruebas de la ocurrencia de ese evento publicados en los diversos informes de inteligencia de fuentes abiertas. Un evento hipotético significa que no existen pruebas del evento y se infieren las razones para demostrar que el evento ha ocurrido.

En esta sección se ha utilizado tres técnicas de modelización de ataques para analizar el caso WhisperGate, donde cada una de las técnicas ofrece una perspectiva diferente sobre un ciberataque. Por ejemplo, el modelo de diamante muestra como un adversario ataca a una víctima dependiendo de dos atributos principales llamados *Infraestructura* y

*Capacidad*. Un ataque tendrá éxito si la *Capacidad* o la *Infraestructura* de la *Victima* son más débiles que las del *Adversario*.

Por otro lado, Cyber Kill Chain entrega los pasos detallados de un ciberataque, proporcionando un modelo eficaz y descriptivo de las operaciones de un atacante facilitando la toma de decisiones de mitigación.

## Incorporación de inteligencia y marcos al análisis de ciberdefensa

La Inteligencia de ciberamenazas (CTI) es un campo en rápido desarrollo que evoluciona en respuesta directa al crecimiento exponencial de los ciberataques. Ha habido muchas contribuciones al campo de la CTI, una de las contribuciones más importantes es el marco MITRE ATT&CK que contiene una lista exhaustiva de tácticas y técnicas del adversario vinculadas a amenazas persistentes avanzadas (APT).[17]

Otro marco desarrollado por MITRE y en colaboración con la Agencia de Seguridad Nacional de EEUU es MITRE D3FEND. Este marco proporciona un modelo para contrarrestar las técnicas ofensivas comunes, enumerando cómo las técnicas defensivas afectan a la capacidad de un actor para tener éxito[18].

La figura siguiente muestra la relación entre el marco ATT&CK y D3FEND.

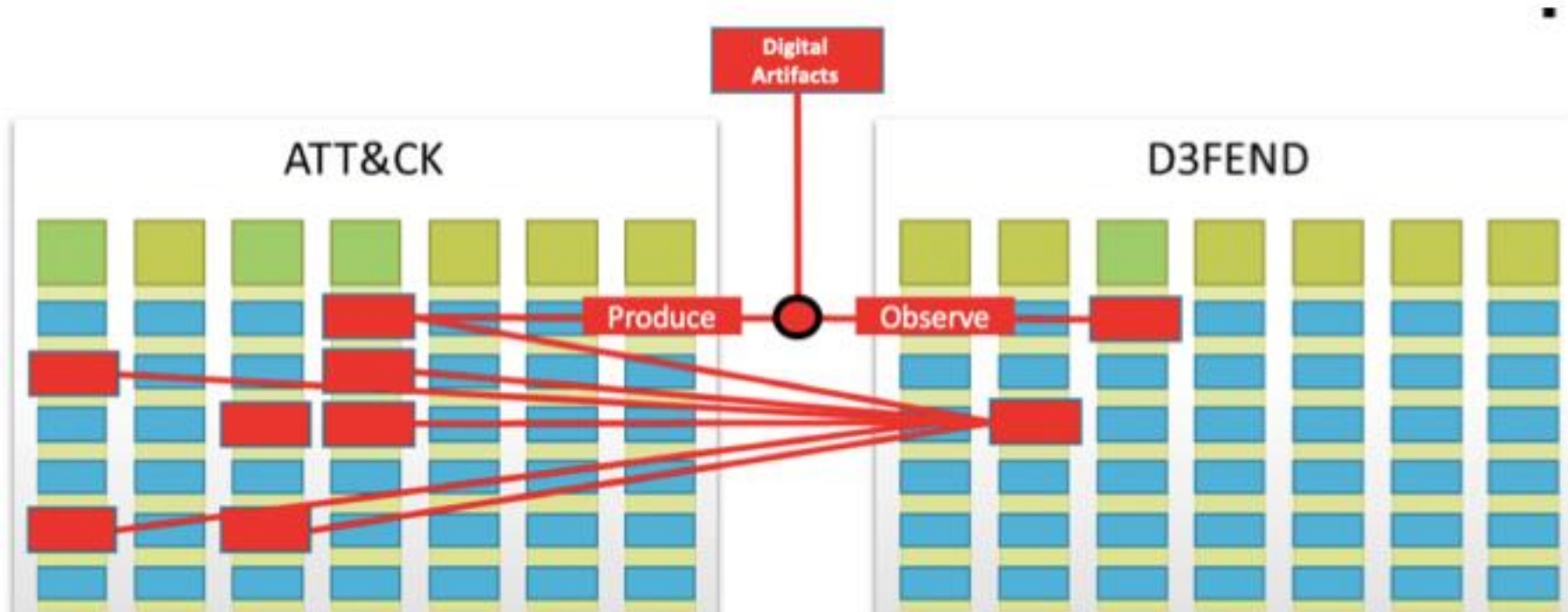

Figura 10: Fuente: MITRE D3FEND https://d3fend.mitre.org/about

---

[17] Kaloroumakis, P. E., & Smith, M. J. (2021). *Toward a knowledge graph of cybersecurity countermeasures*.

[18] Agencia de Seguridad Nacional de EEUU https://www.nsa.gov/Press-Room/Press-Releases-Statements/Press-Release-View/Article/2665993/nsa-funds-development-release-of-d3fend/

En esta sección examinaremos cómo los marcos ATT&CK y D3FEND se pueden utilizar conjuntamente para ayudar a fortalecer el análisis y las contramedidas a un ciberataque y cómo se incorporan ambos marcos a las capacidades de la inteligencia de amenazas.

Retomando el caso WhisperGate analizado en la sección anterior, una de las TTP utilizadas por el grupo DEV-0586 es *T1105 Ingress Tool Transfer* que está categorizada, según el marco ATT&CK, como una táctica de *Comando y Control*. Para establecer las contramedidas de la TTP T1105, es necesario hacer uso del marco D3FEND[19], al buscar la técnica *Ingress Tool Transfer* podemos ver en la figura siguiente las relaciones trazadas por varias contramedidas y esta técnica.

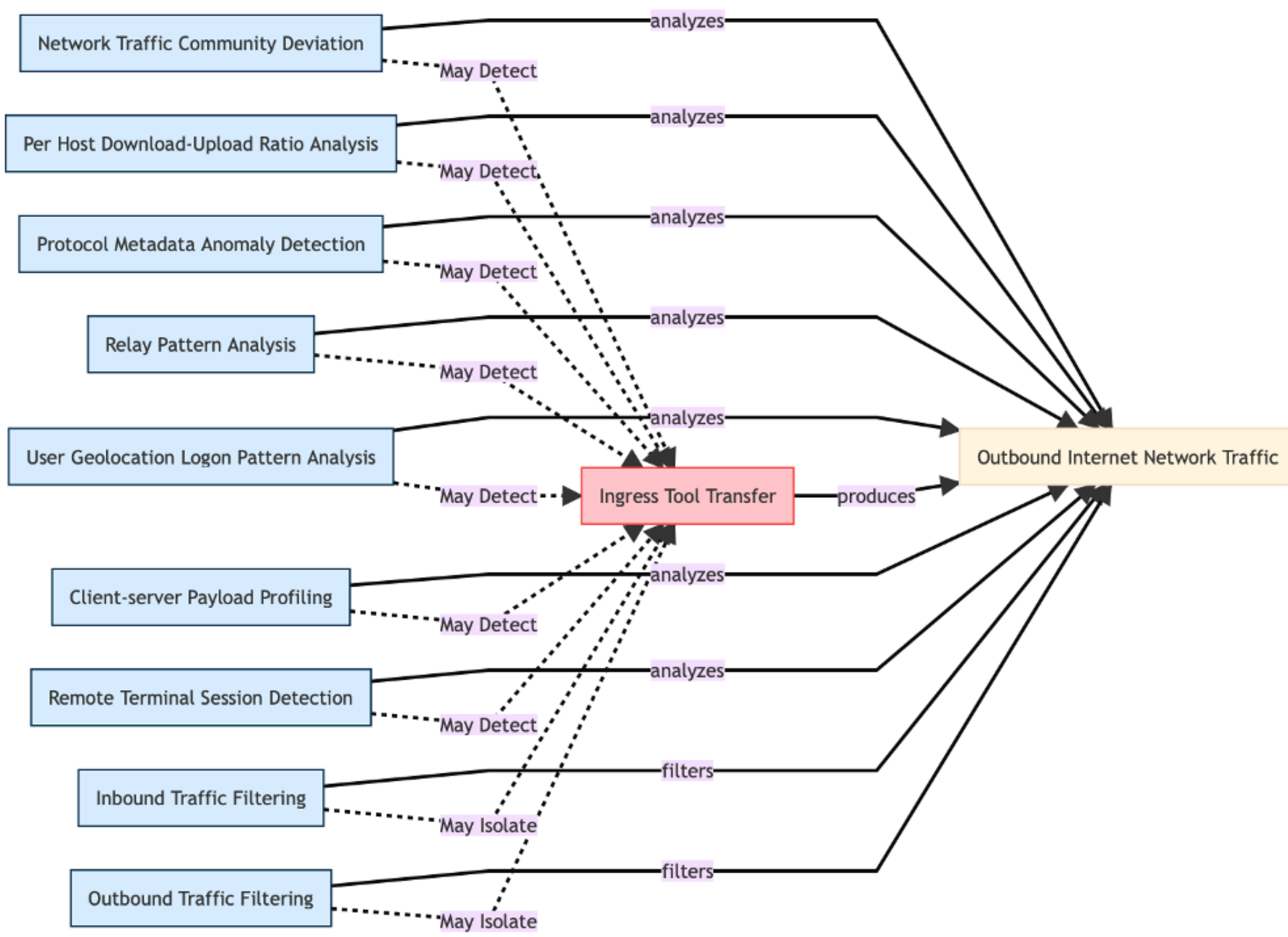

Figura 11: Relaciones inferidas entre ATT&CK y D3FEND

La figura 11 muestra en azul las diversas contramedidas dentro de D3FEND y cómo podrían utilizarse para contrarrestar la técnica *Ingress Tool Transfer*. Con esta información un equipo de ciberdefensa puede empezar a analizar y desplegar contramedidas efectivas.

---

19 Sitio web del proyecto: https://d3fend.mitre.org/

Para este caso de estudio, nos centraremos en la contramedida de *Remote Terminal Session Detection*, mostrada en la Figura 11. El marco D3FEND nos muestra una explicación detallada de las formas de detectar *Remote Terminal Session*, a través de contramedidas Network Traffic Inspection y el Algorithm Analysis Description.

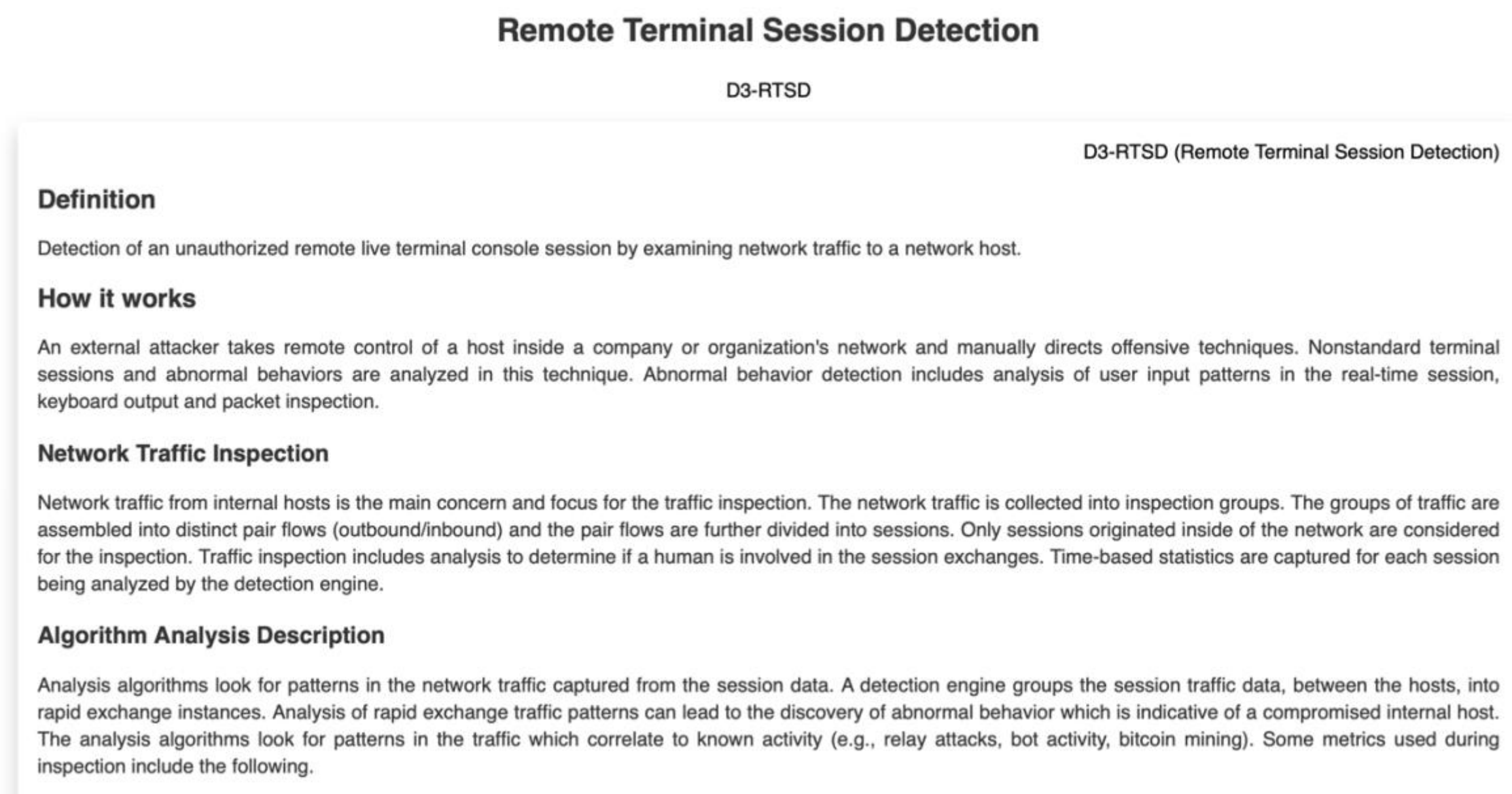

## Remote Terminal Session Detection

D3-RTSD

D3-RTSD (Remote Terminal Session Detection)

### Definition

Detection of an unauthorized remote live terminal console session by examining network traffic to a network host.

### How it works

An external attacker takes remote control of a host inside a company or organization's network and manually directs offensive techniques. Nonstandard terminal sessions and abnormal behaviors are analyzed in this technique. Abnormal behavior detection includes analysis of user input patterns in the real-time session, keyboard output and packet inspection.

#### Network Traffic Inspection

Network traffic from internal hosts is the main concern and focus for the traffic inspection. The network traffic is collected into inspection groups. The groups of traffic are assembled into distinct pair flows (outbound/inbound) and the pair flows are further divided into sessions. Only sessions originated inside of the network are considered for the inspection. Traffic inspection includes analysis to determine if a human is involved in the session exchanges. Time-based statistics are captured for each session being analyzed by the detection engine.

#### Algorithm Analysis Description

Analysis algorithms look for patterns in the network traffic captured from the session data. A detection engine groups the session traffic data, between the hosts, into rapid exchange instances. Analysis of rapid exchange traffic patterns can lead to the discovery of abnormal behavior which is indicative of a compromised internal host. The analysis algorithms look for patterns in the traffic which correlate to known activity (e.g., relay attacks, bot activity, bitcoin mining). Some metrics used during inspection include the following.

Figura 12: Detalle de la contramedida D3FEND Remote Terminal Session Detection

Otra ventaja importante del marco D3FEND es que proporciona mapeos con otras técnicas del marco ATT&CK, donde  la contramedida *Remote Terminal Session Detection* puede ser efectiva, aquí es donde aparece la potencia del uso de ambos marcos combinados, donde una única contramedida puede afectar a docenas de otras técnicas del adversario como muestra la figura 13.

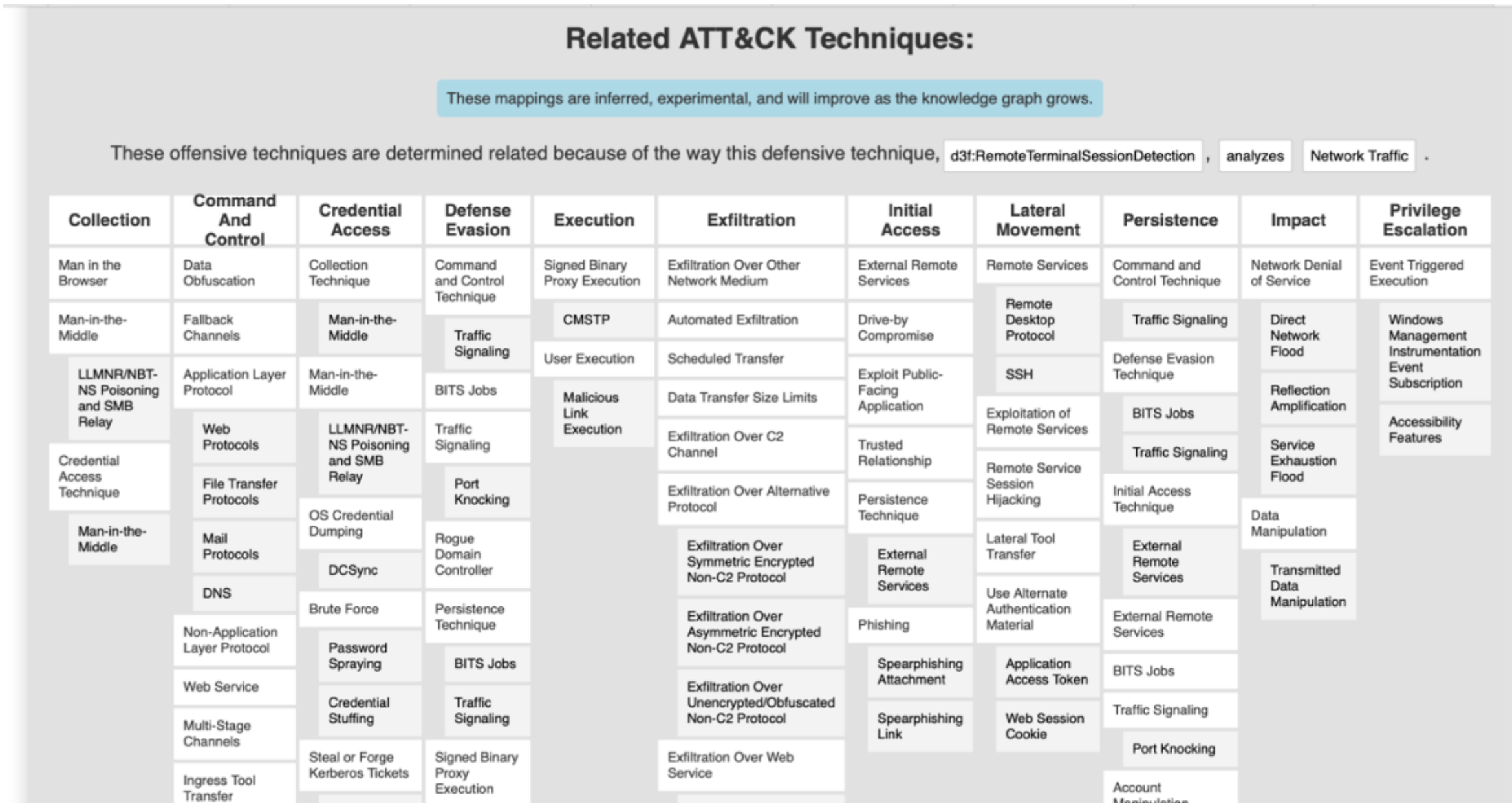

Figura 13: Técnicas de ATT&CK relacionadas con la contramedida D3FEND *Remote Terminal Session Detection*.

El uso combinados de los marcos descritos en este documento muestran la esencia de la Computer Network Defense[20] (CND) enfocada en la inteligencia: basar las decisiones y medidas de seguridad en un conocimiento profundo del adversario[21].

A partir de toda la información obtenida se puede crear una matriz de curso de acción (CoA). Al igual que el marco Cyber Kill Chain representa las distintas fases de un ciberataque desde la perspectiva del adversario, la Matriz CoA es el complemento de las acciones para los equipos de ciberdefensa y permite aplicar medidas para proteger los sistemas en cada una de las fases de la cadena.

La tabla siguiente muestra una matriz de curso de acción utilizando las acciones de detectar, negar, interrumpir, degradar, engañar y destruir[22]. Esta matriz muestra en color azul algunas de las contramedidas sugeridas por el marco D3FEND para las técnicas utilizadas en el ataque WhisperGate.

[20] Computer Network Defense (CND) es un conjunto de procesos y medidas de protección que utilizan redes informáticas para detectar, monitorear, proteger, analizar y defender contra las infiltraciones de red que resultan en la denegación, degradación e interrupciones del servicio/red.
[21] Hutchins, E. M., Cloppert, M. J., & Amin, R. M. (2011). Intelligence-driven computer network defense informed by analysis of adversary campaigns and intrusion kill chains. Leading Issues in Information Warfare & Security Research, 1(1), 80.
[22] U.S. Department of Defense. Joint Publication 3-13 Information Operations, February 2006. URL http://www.dtic.mil/doctrine/new_pubs/jp3_13.pdf.

| FASE | Detectar | Denegar | Interrumpir | Degradar | Engañar | Destruir |
|---|---|---|---|---|---|---|
| Reconocimiento | Web analytics | | | | | |
| Armamento | | | | | | |
| Entrega | **File Analysis** | | | | | |
| Explotación | HIDS | | | | | |
| Instalación | HIDS | **Bootloader Authentication** | **Executable Allowlisting** | | | |
| Mando y Control | **Remote Terminal Session Detection** | | | | | |
| Acción sobre los objetivos | Audit log | | | | Honeypot | |

Tabla 10: Matriz de acciones de Lockheed Martin[23]. Las filas especifican los siete pasos del Cyber Kill Chain, mientras que las columnas especifican las acciones defensivas correspondientes

# Conclusiones y Trabajos Futuros

La defensa informada sobre amenazas (Threat-Informed Defense) se define como *la aplicación sistemática de un profundo conocimiento de la técnica y la tecnología del adversario para mejorar las defensas*.[24]. Gracias al apoyo de la agencia nacional de seguridad de EEUU en conjunto con el Center For Threat Informed Defense de MITRE Corporation, este enfoque está cambiando la forma en la que las organizaciones de todo el mundo mejoran significativamente sus ciberdefensas.

En este trabajo, se propone un enfoque para aplicar la defensa informada sobre amenazas, utilizando distintos marcos para comprender las características de un adversario y construir una defensa basada en la información recolectada.

[23] Ransomware on the rise: An enterprise guide to preventing ransomware attacks. Carbon Black, ebook, February 2017. http://www.bankinfosecu rity.com/whitepapers/ransomware-on-rise-enterprise-guide-to-preventing-ransomware-attacks-w-2760. Accessed Dec 2018

[24] https://www.mitre.org/news/focal-points/threat-informed-defense

En particular, se analiza el ciberataque WhisperGate como caso de estudio. El enfoque propuesto demostró que el grupo DEV-0586 utilizó un conjunto de tácticas y técnicas para infectar a las víctimas y cómo a partir de éstas se pueden proponer medidas defensivas.

La principal contribución de este trabajo es la formalización de una secuencia de pasos sistemáticos que puede utilizarse para la caracterización efectiva de amenazas específicas de un grupo de actividad de interés. Esto se demuestra realizando el levantamiento del perfil de la amenaza, la detección de las tácticas y técnicas utilizadas, a través del marco ATT&CK, y el análisis de la intrusión utilizando los modelos Cyber Kill Chain y Modelo de diamante, para finalmente establecer contramedidas propuestas por el marco D3FEND.

Los trabajos futuros abarcarán la proyección del enfoque a otros ciberataques existentes basados en malware para redes OT[25] que soportan los sistemas ciberfísicos. La seguridad de los sistemas ciberfísicos es de suma importancia, ya que son omnipresentes en las infraestructuras críticas.

[25] La tecnología operativa (OT) consiste en el software y el hardware para controlar los equipos industriales, e incluye los sistemas especializados que se utilizan en los sectores de fabricación, energía, electricidad, minería, etc.

## Bibliografía

Hutchins, E. (2011). *Intelligence-Driven Computer Network Defense Informed by Analysis of Adversary Campaigns and Intrusion Kill Chains .* Lockheed Martin Corporation. Obtenido de https://www.lockheedmartin.com/content/dam/lockheed-martin/rms/documents/cyber/LM-White-Paper-Intel-Driven-Defense.pdf

Castro, S. (2021). Towards the Development of a Rationalist Cyber conflict Theory. *The Cyber Defense Review 6(1)*, 35-62.

Michael Assante, R. L. (2015). *The industrial control system cyber kill chain.* InfoSec Reading Room,1.

*MIT Technology Review*. (21 de Enero de 2022). Obtenido de MIT Technology Review: https://www.technologyreview.com/2022/01/21/1043980/how-a-russian-cyberwar-in-ukraine-could-ripple-out-globally/

Use, C. D. (2016). Analysis of the cyber attack on the Ukrainian power grid. *Electricity Information Sharing and Analysis Center (E-ISAC), 388*, 1-29.

CHALFANT, M. (18 de 2 de 2022). *White House says Russia behind cyberattack on banks, ministry in Ukraine*. Obtenido de https://thehill.com/policy/cybersecurity/594947-white-house-says-russia-behind-cyberattack-on-banks-in-ukraine

*helpnet security*. (12 de 4 de 2022). Obtenido de Helpnet Security: https://www.helpnetsecurity.com/2022/04/12/sandworm-ukraine/

*Lisa Institute*. (3 de 3 de 2020). Obtenido de LISA Institute: https://www.lisainstitute.com/blogs/blog/ciberinteligencia-que-es-y-para-que-sirve

*Business News Daily*. (17 de 2 de 2022). Obtenido de Business News Daily: What is Cyber Threat Intelligence, and Why Do You Need It?" Business News Daily. Available at https://www.businessnewsdaily.com/11141-cyber-threat- intelligence.html

Caltagirone, S. P. (2013). The diamond model of intrusion analysis. *Center For Cyber Intelligence Analysis and Threat Research Hanover Md*.

*Cybersecurity & Infrastructure Security Agency*. (28 de 4 de 2022). Obtenido de Cybersecurity & Infrastructure Security Agency: https://www.cisa.gov/uscert/ncas/alerts/aa22-057a